\documentclass[twocolumn,final,5p]{mystyle}

\usepackage{tabularx}

\usepackage{times}
\usepackage{graphicx}
\usepackage{array}
\usepackage{amsmath}
\usepackage{amssymb}
\usepackage{algorithm}
\usepackage{algorithmic}
\usepackage{url}
\usepackage{multirow}

\usepackage{subfigure}

\newcolumntype{V}{>{$\vcenter\bgroup\hbox\bgroup}c<{\egroup\egroup$}}
\def\Hline{\noalign{\hrule height 4\arrayrulewidth}}

\graphicspath{{../../}}

\begin{document}

\begin{frontmatter}

\title{Fairer Citation-Based Metrics}
\author[DEAK]{Ognjen Arandjelovi\'c}

\address[DEAK]{Centre for Pattern Recognition and Data Analytics (PRaDA)\\Deakin University, Australia\\
E-mail: \texttt{ognjen.arandjelovic@gmail.com}\\Web: \texttt{http://mi.eng.cam.ac.uk/$\sim$oa214} }

\begin{abstract}
I describe a simple modification which can be applied to any citation count-based index (e.g.\ Hirsch's $h$-index) quantifying a researcher's publication output. The key idea behind the proposed approach is that the merit for the citations of a paper should be distributed amongst its authors according to their relative contributions. In addition to producing inherently fairer metrics I show that the proposed modification has the potential to partially normalize for the unfair effects of honorary authorship and thus discourage this practice.
\end{abstract}


\end{frontmatter}

\section{Introduction}
The concept of quantification is intrinsic to the scientific method. Considering the central and pervasive role that quantification plays in science, it should come as no surprise that the magnifying glass would be turned back on itself and that scientists would want to quantify aspects of their own work. In particular in this paper I am interested in considering various indexes which have become commonplace metrics of a researcher's output.

Broadly speaking, the intended purpose of indexes discussed herein is to ``quantify the cumulative impact and relevance of an individual's scientific research output'' to quote Hirsch, the author of one of the most widely used indexes~\cite{Hirs2005}. What is more they aim to achieve this quantification using citation statistics of the individual's publications as the observable input measurements. This very idea has produced much controversy~\cite{Yong2014,Bati2006,Meho2007}. I too argue that the subjective understanding of what `impact' means in this context inherently makes the very aim of its objective quantification a non-scientific proposition. Considering the lack of an objective basis, the ground truth if you will, for assessing the performance of a particular index, unlike different previous authors  (e.g.\ see $h$-index~\cite{Hirs2005}, $e$-index~\cite{Zhang2009}, $g$-index~\cite{Eggh2006}, $z$-index~\cite{PeteSucc2013}, $i10$-index~\cite{Goog2011}) who have described and argued in favour of different indexes~\cite{BornMutzDani2010,Vaid2005} in this paper I do not propose a novel index \textit{per se}. Rather, accepting the pragmatic standpoint that for better or worse citation indexes \emph{are} being increasingly used in academia~\cite{Meho2007}, I show how a simple modification, applicable to any citation count-based index, can make it \textit{ipso facto} fairer.

\section{Contribution-weighted citations}
As the starting point to motivate the key idea, contemplate the following thought experiment and the question which naturally emerges from it. Consider a particular publication and two alternative scenarios: in one scenario the entire work is performed by a single author, in the other by two or more authors. The question I ask is: Is the contribution to the paper's impact of the sole author in the former scenario equal to the contributions of each of the authors of the latter scenario? Given that the totality of the work is the same and that in the latter case it is produced by a joint effort, it seems clear that the answer is no. What is more in the latter scenario the claim by each of the authors to the total impact of the work should not be equal but portioned according to the authors' relative contributions. Therefore I propose the following. Let us express a specific citation-based index as a function $f(\text{citations}_1, \ldots, \text{citations}_n)$ where $\text{citations}_i$ is the number of citations of a person's $i$-th publication (of $n$ in total). I argue that regardless of the index used, i.e.\ regardless of the form of $f$, a fairer quantification using the same baseline idea can be achieved by evaluating:
\begin{align}
   \stackrel{o}{f}&(\text{citations}_1, \ldots, \text{citations}_n) = \notag\\
      &f(\text{auth\_rank}_1^{-1} \times \text{citations}_1, \ldots, \text{auth\_rank}_n^{-1} \times \text{citations}_n),
   \label{e:oix}
\end{align}
where $\text{auth\_rank}_i$ is the rank of the researcher in the list of $i$-th paper's authors.

The expression in Eq.~\eqref{e:oix} exploits the observation that the order of individuals in the list of a paper's authors conveys information about their relative contributions: the first author contributed at least as much as the second, the second at least as much as the third, and so on. This allows us to derive the upper bound of the relative contribution of the $i$-th author as $\text{auth\_rank}_i^{-1}$. It is simple to see that this upper bound is achieved when the first $i$ authors contribute equal amounts and the remaining authors nothing at all.

It is important to recognize the crucial difference in what I propose and the previous work on research output quantification. In particular I am referring to the nature of the sole assumption I make: that the ordering of authors reflects their relative contributions. Its validity is virtually ensured by the competing interests of authors; for one author to be promoted to a higher rank in the list of the paper's authors, another one must be demoted. This is in stark contrast to previous ideas which align the interests of all authors of a single paper and thus provide incentive to researchers to act in ways other than in ``the best spirit'' of academic publishing (e.g.\ by adding to the list of authors individuals who had not contributed to the work -- I will come back to this shortly).

\section{Analysis and discussion}
\begin{figure}[htb]
  \centering
  \includegraphics[width=0.35\textwidth]{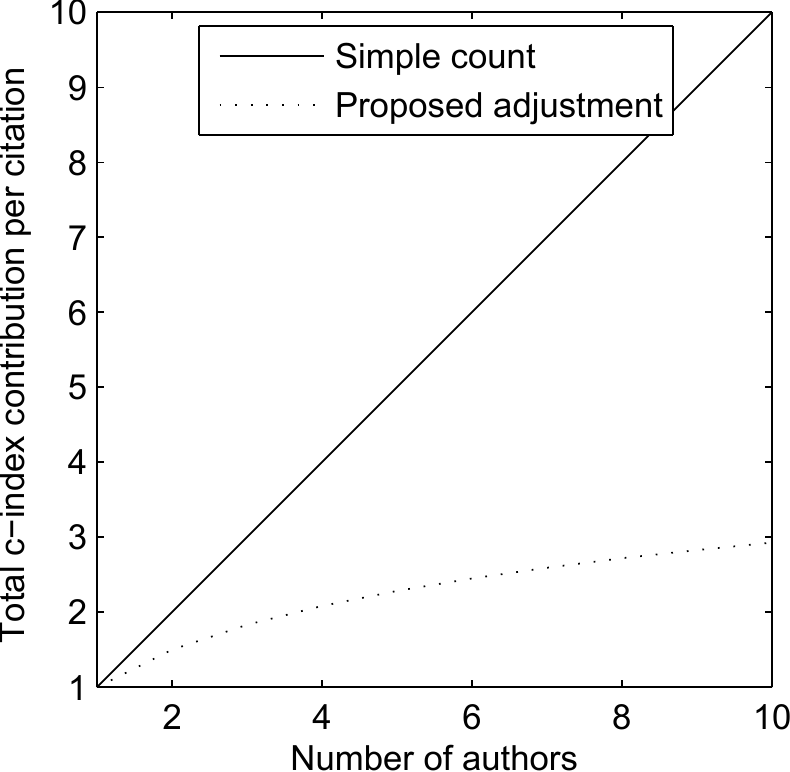}
  \caption{ Total $c$-index impact per citation as a function of the number of paper's authors. }
  \label{f:totalCredit}
\end{figure}

Recall that one of the key ideas motivating the proposed modification is that the total merit for the paper's impact should be unaffected by the number of researchers that authored the work. Consider the simple citation count quantification of output, the $c$-index for short. If a particular paper was authored by $n$ authors and cited $c$ times, the totality of the merit for the paper's impact is $nc$ since the citation count $c$ contributes to all of the authors' $c$-index. Clearly, this is a linear function of $n$. If the proposed modification is introduced, the total merit becomes $\sum_{i=1}^n i^{-1}$. While this is still a function of $n$ (the ideal characteristic would be a horizontal line with the ordinate value of 1), the growth is very much supralinear, as illustrated in Fig.~\ref{f:totalCredit}. In addition to the fundamental argument laid out previously, this is important because it disincentivizes dishonest addition of non-contributing persons to the list of a paper's authors.  Without the proposed modification, the incentive is high because all individuals involved stand to gain benefit, e.g.\ the person added as an author gets additional merit from all the citations of the paper while the actual authors of the paper gain by the expected reciprocal behaviour (i.e.\ by being themselves added as authors to papers that they have not contributed to)~\cite{FlanCareFontPhil+1998}. While the situation remains a positive sum game, with the proposed modification the incentive for such behaviour is much reduced by the quickly diminishing benefit to lower ranking authors. This remains the case when the modification is applied to other indexes too. For example, consider the $h$-index. For a paper to increase a researcher's $h$-index $h$, it is necessary (but not sufficient) that it receives at least $h$ citations; in contrast, when the proposed modification is applied, the required number of citations becomes $h \times \text{auth\_rank}$.

\begin{figure*}[htb]
  \centering
  \subfigure[]{\includegraphics[width=0.30\textwidth]{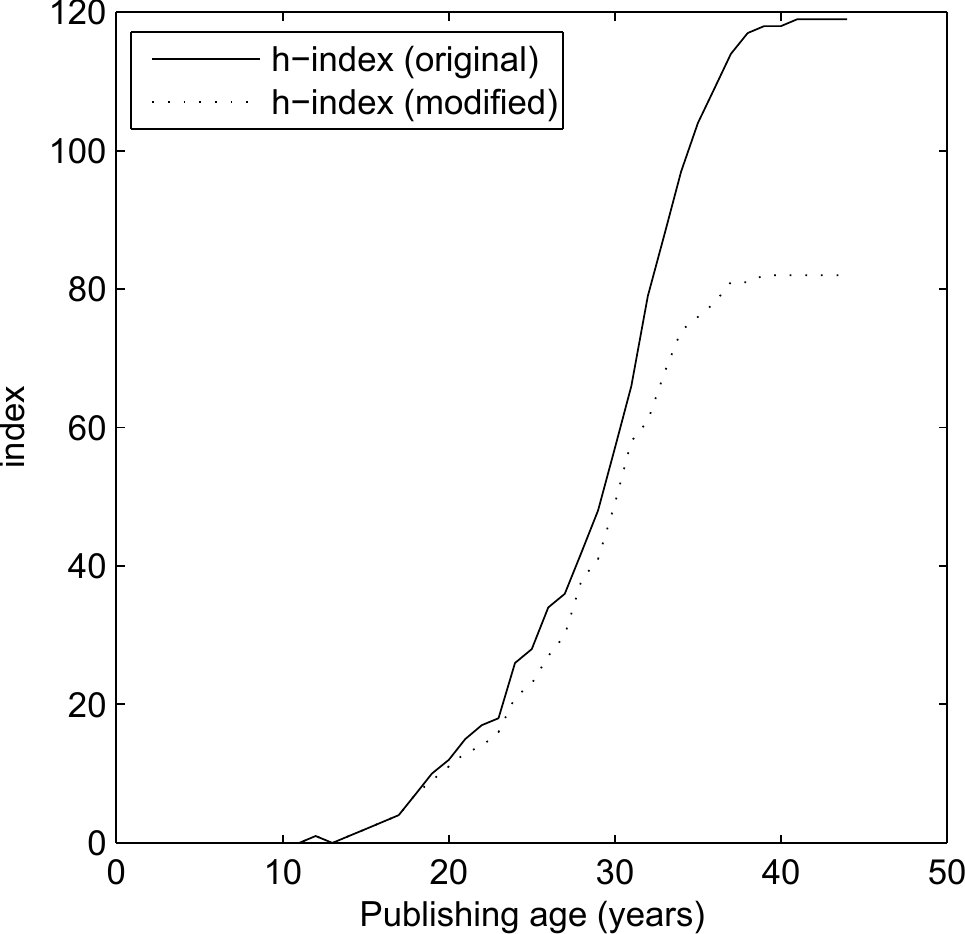}\label{f:hindex}}~~~~~~~~~~~~
  \subfigure[]{\includegraphics[width=0.30\textwidth]{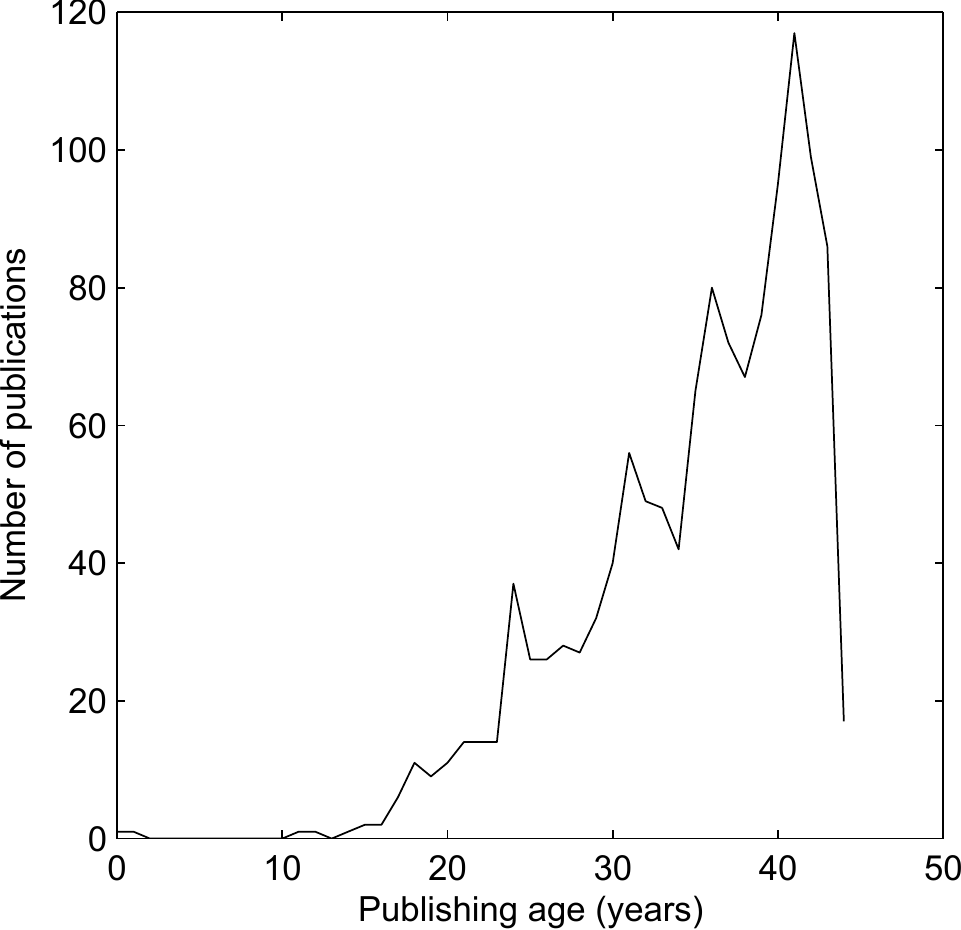}\label{f:numPapers}}~~~~~~~~~~~~
  \subfigure[]{\includegraphics[width=0.30\textwidth]{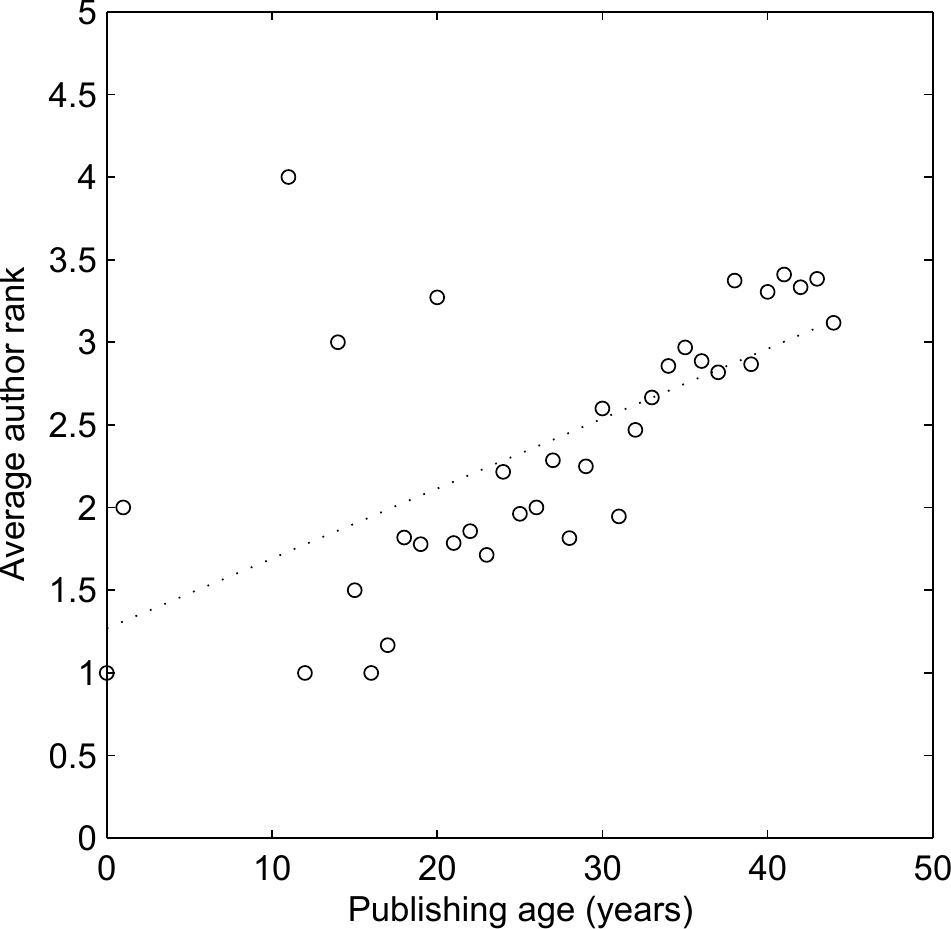}\label{f:authorRank}}
  \caption{ An example of a highly cited researcher at a leading university: the person's (a) $h$-index, in its original form and with the modification proposed herein, computed using only papers published up to a certain year (abscissa value), (b) per annum publishing rate, and (c) the average rank in the list of authors of authored papers.  }
\end{figure*}

Let us consider in some more detail how the proposed modification affects Hirsch's $h$-index. Using a simple publishing model in which a researcher publishes $p$ papers per year, each of which gets cited $c$ times every subsequent year, Hirsch showed that the corresponding $h$-index is a linear function of the researcher's publishing age $n$:
\begin{align}
  h=\frac{c}{1+c/p}~n.
\end{align}
Using the same publishing model a similar derivation can be used to show that the relationship between a researcher's $h$-index calculated using only those papers published in the first $y$ years (but all citations to date), and $y$ is also linear. However, I find that this is seldom the case in practice. This may not be particularly surprising considering the limitations of the simple publishing model used; however, what does need further examination is the observation that nearly universally the actual relationship is superlinear. An example, using a successful researcher at a leading university, is shown in Fig.~\ref{f:hindex} (solid line). The significantly superlinear increase is readily apparent (the final leveling off being caused by the limited time that the recent publications have been available for citation), with the $h$-index increasing approximately six-fold in the second half of the researcher's career. There may be numerous factors involved in this: one's growing academic reputation increases the awareness of the person's research and with it the overall citation rate (creating a positive feedback loop), in some fields accumulated experience plays a role in increasing the quality of published work, and so on. However, further analysis suggests a more worrying dominant factor. The plot in Fig.~\ref{f:numPapers} shows the number of papers published per year by the same researcher. Not only is the publication rate not constant across the researcher's publishing career, as assumed in Hirsch's simple model, but it is steadily increasing. It is remarkable to notice that the number of papers authored by this researcher in the peak publishing year is 117 -- this is a rate of one paper every three days. I would suggest that it is most unlikely that a single individual could have contributed to 117 publications in one year to a sufficient degree to meet the authorship threshold for all of them. Further insight is provided by the data show in in Fig.~\ref{f:authorRank}, which shows the average rank of the researcher in the list of authors across all authored papers for each year of the researcher's career. Here too the trend is clearly evident: the researcher's has steadily been moving down the list of authors, often publishing papers as the leading author during the first 15 years of the career, and most often as the third author in recent years. In and of itself this is not a problem; indeed this trend is typical in most fields of research and can be a reflection of a shift in the nature of the person's contributions. Nevertheless, taken in the context of the previously presented data, namely the extraordinary publication rate and the associated rapid increase in the researcher's $h$-index, the totality of evidence suggests an increasing amount of so-called honorary authorship -- the practice of a senior research member (such as the head of a laboratory or a research group) being included as an author to all publications produced by the lab without actually contributing to the work itself~\cite{Hold2006,FlanCareFontPhil+1998}. Such practice contravenes the norms of `best academic practice'; to quote the uniform requirements of the International Committee for Medical Journal Editors for manuscripts submitted to biomedical journals~\cite{ICMJE}:
\begin{quote}
  ``Authorship credit should be based on (1) substantial contributions to conception and design, acquisition of data, or analysis and interpretation of data; (2) drafting the article or revising it critically for important intellectual content; and (3) final approval of the version to be published. Authors should meet conditions 1, 2, and~3.''
\end{quote}

Now let us consider the temporal behaviour of the $h$-index when the modification I propose is applied. Adopting the simple publication model of Hirsch, it is easy to see that if the authorship rank of a particular researcher is the same in all publications, the modified $h$-index $\stackrel{o}{h}$ also grows linearly with $y$, albeit at a rate slower by a factor of $\text{auth\_rank}$ (clearly if $\text{auth\_rank}=1$ then $\stackrel{o}{h}$ becomes equal to $h$). However, this is a rather unrealistic assumption; in most cases a researcher publishes as the leading author in the early stages of the career, and over time with the increase in seniority contributes to research in a more supervisorial fashion (a valuable contribution entirely in accordance with authorship credit requirements; not to be confused with honorary authorship which by definition is neither). This shift has two effects on $\stackrel{o}{h}$. The first of these acts so as to reduce $\stackrel{o}{h}$ because of the weighting of citations by $\text{auth\_rank}^{-1}$. On the other hand, the more abstracted nature of contributions typical for senior researchers allows a person to contribute to a greater number of papers thereby acting so as to increase $\stackrel{o}{h}$. Considering that the portioning of merit described in Eq.~\eqref{e:oix} allocates the upper bound of possible merit to each author, in most cases it can be expected that the latter of the two forces would prevail and that $\stackrel{o}{h}$ would exhibit superlinear growth. The example given in Fig.~\ref{f:hindex} (dotted line) is consistent with this prediction.

Working on the assumption of a linear relationship between $h$ and $n$ and writing $h \approx mn$, Hirsch argued that the coefficient $m$ can be used to compare researchers with different publishing ages. Hirsch finds that $m \approx 1$ characterizes ``successful scientists'', $m \approx 2$ ``outstanding scientists,
likely to be found only at the top universities or major research laboratories'', and $m \approx 3$ or higher ``truly unique individuals''. In Table~\ref{t:topCS} I summarize several relevant statistics for computer scientists with a Google Scholar computed $h$-index of at least 110 at the time of writing of this article. There are 11 names on this list, with the average $h$-index of 118 ($\sigma_h=9.1$), average citation count of 77,898 ($\sigma_c=15,346$), and the average value of the $m$ parameter of 3.5 ($\sigma_m=0.64$). The same table also shows the corresponding statistics when the modification proposed in this paper is applied: the average $\stackrel{o}{h}$-index becomes 69 ($\sigma_{\stackrel{o}{h}}=12.9$), the average adjusted citation count 31,830 ($\sigma_{\stackrel{o}{c}}=17,295$), and the average value of $\stackrel{o}{m}$ 2.0 ($\sigma_{\stackrel{o}{m}}=0.50$). These are reductions of respectively 42\%, 60\%, and 41\%.

\begin{table*}[htb]
  \caption{Examples of citation-based impact metrics for computer scientists with an $h$-index of at least 110, without and with the proposed modification.}
  \vspace{8pt}
  \renewcommand{\arraystretch}{1.2}
  \centering
  \begin{tabular}{lcccccc}
  \Hline
     Researcher      & $h$-index & $\stackrel{o}{h}$-index & $c$-index & $\stackrel{o}{c}$-index & Rate $m$ & Rate $\stackrel{o}{m}$\\
     \hline
     A K Jain        & 142     & 101      & 108,828   & 59,721 & 3.5 & 2.5\\
     T Sejnowski     & 123     &  78      &  79,159   & 29,436 & 2.7 & 1.7\\
     S Shenker       & 122     &  60      &  83,516   & 18,519 & 3.5 & 1.7\\
     H Garcia-Molina & 120     &  65      &  61,023   & 18,753 & 3.2 & 1.8\\
     J Han           & 120     &  68      &  87,977   & 50,982 & 3.7 & 2.1\\
     T Poggio        & 114     &  69      &  68,072   & 29,292 & 2.6 & 1.6\\
     D Haussler      & 113     &  54      &  85,684   & 13,040 & 2.5 & 1.2\\
     S Thrun         & 113     &  64      &  54,788   & 18,419 & 4.5 & 2.6\\
     M I Jordan      & 113     &  71      &  73,541   & 32,266 & 4.0 & 2.5\\
     I Foster        & 111     &  77      &  88,714   & 60,151 & 4.0 & 2.8\\
     A Zisserman     & 110     &  57      &  65,580   & 19,551 & 3.8 & 2.0\\
     \Hline
  \end{tabular}
  \label{t:topCS}
  \vspace{10pt}
\end{table*}

\section{Summary and conclusions}
In this paper I described a general modification which can be applied to any citation-based metric of an individual's research output. The key idea was to distribute the merit for the citations of a paper amongst its authors according to their relative contributions inferred from the authorship order. I argued that the validity of this approach is ensured by the competing interests of different authors. Using both theoretical arguments and empirical examples, I showed that the proposed modification has the potential to partially normalize for the unfair effects of honorary authorship and thus discourage this practice. Lastly, it should be noted that the proposed modification ceases to be useful when a researcher has publications in venues which use alphabetical ordering of authors. Today this is rare -- a recent survey estimates that this practice is maintained by less than 4\% of academic journals, with a decreasing trend~\cite{Walt2012}.

\bibliographystyle{unsrt}
\bibliography{../../../../my_bibliography}
\end{document}